\documentclass[fleqn,10pt]{wlscirep}
\usepackage[utf8]{inputenc}
\usepackage[T1]{fontenc}
\title{A microwave scattering spectral method to detect the nanomechanical vibrations embedded in a superconducting qubit}

\author[1]{H. Y. Gao}
\author[1,*]{L. F. Wei}
\affil[1]{Information Quantum Technology Laboratory, International Cooperation Research Center of China Communication and Sensor Networks for Modern Transportation, School of Information Science and Technology, Southwest Jiaotong University, Chengdu 610031, China}

\affil[*]{lfwei@swjtu.edu.cn}


\begin{abstract}
Nanomechanical resonators (NMRs), as the quantum mechanical sensing probers, have played the important roles for various high-precision quantum measurements. Differing from the previous emission spectral probes (i.e., the NMR modified the atomic emission), in this paper we propose an alternative approach, i.e., by probing the scattering spectra of the quantum mechanical prober coupled to the driving microwaves, to characterize the physical features of the NMR embedded in a rf-SQUID based superconducting qubit. It is shown that, from the observed specifical frequency points in the spectra, i.e., either the dips or the peaks, the vibrational features (i.e., they are classical vibration or quantum mechanical one) and the physical parameters (typically such as the vibrational frequency and displacements) of the NMR can be determined effectively. The proposal is feasible with the current technique and should be useful to design the desired NMRs for various quantum metrological applications.
\end{abstract}
\begin{document}
\flushbottom
\maketitle
\thispagestyle{empty}
\section*{Introduction}

In recent years, nanomechanical resonators (NMRs) have been recently highlighted for various precise measurements, as their mechanical vibrations can reach very high frequency such as up to the GigaHertz (see, e.g.,~\cite{NMRF}). This makes them be directly utilized as the electronic devices for radio communications and various precise measurements such as for sensing masses, weak forces, and charge, etc.~\cite{physrep,EPL}. In fact, the ability of the resonator to detect the physical quantities is closely related to its resonant frequency. For example, the mass-loaded sensitivity can be written as $s=\delta f_n/\delta m=f_n/2m$~\cite{mass,EEE2019,2012J,SSS2013}, if the resonator with the eigenfrequency $f_n$ and mass $m$ is added by a mass $\delta m$, which leads to the frequency shift $\delta f_n$. This implies that the higher frequency of the resonator corresponds to the stronger ability to detect the smaller masses, and also the vibrational frequency of the NMR should be precisely calibrated beforehand.\par
Physically, a classical oscillator has a well-defined amplitude of motion; while, for the quantum oscillator, its displacement $z$ is related to the vibrational quantum state with the quantum fluctuations of the momentum and displacement. In fact, by various cooling techniques, the vibrations of the NMRs can be cooled to the approaching quantum ground state, starting from a thermal state~\cite{T2010}. As a consequence, the NMR with the GigaHertz vibrational frequency can be served as the coherent quantum devices to test the fundamental principles in quantum mechanics (e.g., the Heisenberg uncertainty relation, quantum superposition, and macroscopic quantum, etc.) and also generate various hybrid quantum systems (e.g., coupling it to the superconducting circuits~\cite{N2018,X17}, semiconducting quantum dots~\cite{17,N16}, and NV centers~\cite{Z2017,xiao2017}, etc.\cite{P2018}) for implementing the desired quantum metrologies and quantum information processings~\cite{14,15,16}. Therefore, characterizing the physical features (i.e., the vibration is classical or quantum mechanical) and further measuring the relevant physical parameters (typically such as the vibrational frequency and displacement) of the vibration of the NMR are particularly important.

Basically, compared with the calibrations of the physical parameters of the classical mechanical vibration reviewed below, the precise measurement of the vibrational displacement of a quantized high frequency harmonic resonator (HO) is still an open problem. This is because that, besides the various background noises, the internal quantum fluctuations of the vibration plays a key role. For example, the amplitude $A_0$ of a quantized HO at vibrational ground state $|0\rangle$ is directly determined by the quantum fluctuation of the measured vibrational displacement, i.e., $A_0=\varrho_z/\sqrt{2}$ with $\varrho_z=\sqrt{\langle 0|\hat{z}^2|0\rangle-\langle 0|\hat{z}|0\rangle^2}$. Also, the sensitivity of the frequency measurement of a quantized HO is also limited by the uncertainty relation between the quantized vibrational energy and the lifetime of the operated energy stationary state. Up to our knowledge, a few methods have been demonstrated to detect the displacement of the quantized vibration of the NMR at low temperature. For example, the vibrational frequency of the quantized NMR, which is embedded in a superconducting transmission line resonator (STLR), can be measured by using a microwave interferometer configuration~\cite{C2008}. The motion of the quantized vibrations of the NMR can be measured and control by probing the optical-mechanical effects in the cavity optomechanical system~\cite{A2014}. Typically, in Ref.~\cite{wei} we proposed an effective approach to probe the tiny motion of the NMR by coupling it to a half-wavelength STLR, mediated by a SQUID-based qubit. In that configuration, the quantized mechanical vibration of the NMR modifies the energy structure of the qubit and thus its spontaneous mission spectrum, which can be indirectly detected by the spectral measurement of the STLR. Interestingly, the vibrational features, i.e., the vibration is quantum mechanical or classical, can be identified by observing the modifications in the spectrum spontaneously emitted from the SQUID-based qubit. Alternatively, in the present work we propose an active approach (rather than the passive one in Ref.~\cite{wei}) to probe the mechanical vibrations of the NMR by measuring the transmission spectra of the travelling microwave along a one-dimensional transmission line. The detected NMR in the present configuration is embedded in a rf-SQUID-based superconducting qubit. As a consequence, the vibrational frequency and displacement of the NMR can be estimated by observing a few specific frequency points in the transmitted spectra of driven travelling microwave scattered by the qubit. Furthermore, we show that the proposal works also for the alternative measurement configuration, i.e., the scatter of the microwave is replaced by a quarter-wavelength STLR (with a sufficiently-high quality factor), which is inductively coupled to the NMR via the qubit. Due to the use of the STLR, the electromagnetically induced transparency-like effects in microwave band are modified and thus the relevant parameter estimations could be more conveniently achieved with the more observable data. Importantly, a recent experiment~\cite{NN2019} had demonstrated the resolution detection of the energy levels of a NMR by using the scattering spectral measurements of the travelling microwaves.

 We propose a spectral approach, by probing the transmitted and phase shift spectra of the travelling wave scattered directly by the qubit-NMR device, to estimate the physical parameters of the NMR embedded in a rf-SQUID qubit. Then we treat the problem with a more complicated one, i.e., the STLR-qubit-NMR system, and demonstrate the corresponding spectral measurements of the NMRs. One can see that, with these spectra the desired physical parameters can be more easily estimated by using the more observable data, due to the microwave electromagnetically induced transparency-like effects. Finally, in order to the completeness, we review how the physical parameters of a classical mechanical resonator were measured and give he derivations of the relevant Hamiltonians.
\section*{Results}

\subsection*{Measuring the frequency and amplitude with a mircowave driven qubit-NMR system}
The NMR considered here is generated by the mechanical vibration of the part of a flux-biased rf-SQUID loop, which generates a qubit encoded by the two lowest energy eigenstates of the loop. Another magnetic field ${\bf B}_0$ can be applied along the loop plane to provide a restoring force for generating a mechanical vibration of the part of the loop, i.e., the NMR oscillates along the direction perpendicular to the loop plane. The significantly weak vibration of the NMR along the direction parallel to the loop plane~\cite{A2008,xu} can be omitted for the simplicity. Certainly, the other configurations can also be utilized to realize the qubit-NMR couplings, see, e.g., in Refs.~\cite{EB2006,JZ2009}.

\begin{figure}[ht]
  \centering
  \includegraphics[width=8cm]{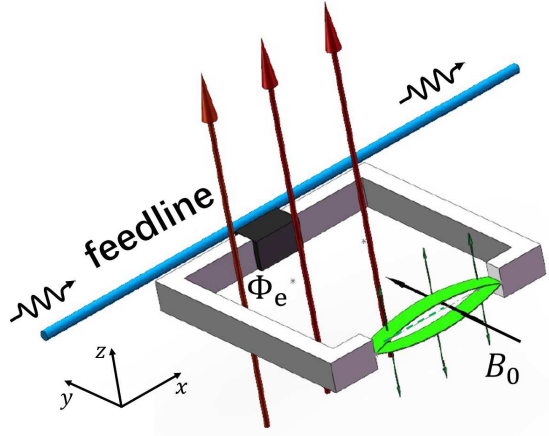}\\
  \caption{Travelling microwave scattered by a qubit-NMR device for calibrating the physical parameters of the NMR embedded in a rf-SQUID loop. Here, the black part represents the Josephson junction, the red arrow perpendicular to the loop indicates the direction of the biased external flux $\Phi_{e}$. The black arrow perpendicular to NMR indicates the direction of another external magnetic field $\bf{B}_{0}$, which excites the mechanical vibration of the NMR (green) along the $z$ direction.}
  \label{Fig1}
\end{figure}

The qubit can served as a probe to measure the physical features of the coupled NMR by using the scattering measurements of the travelling microwave. Without loss of the generality, let us consider the specifical configuration shown schematically in Fig.~\ref{Fig1}. The Hamiltonian of the system can be expressed as ($\hbar=1$)
\begin{equation}\label{eq5}
\hat{H}_1=\hat{H}_f+\hat{H}_{fs}+\hat{H}_{s},
\end{equation}
where
\begin{equation}\label{6}
\hat{H}_f=\int dx[\hat{c}_{R}^{\dagger}(x)(-iv_{g}\frac{\partial}{\partial x})\hat{c}_{R}(x)
+\hat{c}_{L}^{\dagger}(x) (iv_{g}\frac{\partial}{\partial x} )\hat{c}_{L}(x)]
\end{equation}
describes the quantized traveling microwave (with the group speed \(v_g\)) transporting along the feedline~\cite{Sh2009}. $\hat{c}_{L/R}(x)$ and $\hat{c}^\dagger_{L/R}(x)$ are the $x$-dependent annihilation and creation operators of the left/right-moving microwave photons, respectively.
Next,
\begin{equation}\label{7}
\hat{H}_{fs}=\int dx V_1\delta(x)\sum_{j=L,R}[\hat{c}_{j}^{\dagger}(x)\hat{\sigma}_{-}+\hat{\sigma}^{\dagger}\hat{c}_{j}(x)],\\
\end{equation}
describes the interaction (with the strength $V_1$) between the microwave photons transporting along the feedline and the qubit~\cite{Sh2009}. \(\hat{\sigma}^\dagger\) and \(\hat{\sigma}_-\) are the Pauli operators of the qubit. As the wavelength of the microwave is significantly longer than the scale of the rf-SQUID loop, the interaction between them can be treated as a $\delta$-function, taking place at $x=0$. Thirdly, the Hamiltonian $\hat{H}_{s}$, describing the NMR and its coupling to the qubit, takes the forms depending on the specifical features of the vibration of the NMR~\cite{wei}.

Theoretically, the microwave scattering features of the qubit-NMR system can be calculated by solving the time-dependent Schr\"odinger equation:
\(i\partial|\psi(t)\rangle/\partial t=\hat{H}_1|\psi(t)\rangle\),
for the usual elastic scattering, the problem becomes to solve the stationary Schr\"odinger equation:
\begin{equation}\label{8}
\hat{H}_1|\psi\rangle=\omega|\psi\rangle,
\end{equation}
by letting \(|\psi(t)\rangle=e^{-i\omega t}|\psi\rangle\). Here, \(\omega=v_gk\) (\(k\) being the wave vector of the applied microwave transporting along the feedline.

\subsubsection*{Measuring the eigenfrequency of the qubit}
First, if the NMR is absent, i.e., the magnetic field $\bf{B}_0$ is not applied, then $\hat{H}_s$ in Eq.~(\ref{eq5}) is nothing but the Hamiltonian of the rf-SUID-based qubit:
\begin{equation}\label{9}
\hat{H}_q=\omega_{0}|1\rangle\langle 1|,
\end{equation}
with $\omega_0$ being the energy of the qubit's excited state $|1\rangle$. In this case, the generic solution to the equation (\ref{8}) can be expressed as
\begin{eqnarray}\label{10}
|\psi_{0}\rangle={\int dx[\phi _{R}(x)\hat{c}_{R}^{\dagger }(x)+\phi_{L}(x)\hat{c}_{L}^{\dagger}(x)]|\phi_{0}\rangle+A_{0}\hat{\sigma}^{\dagger}|\phi_{0}\rangle},
\end{eqnarray}
where \(|\phi_{0}\rangle=|0,0\rangle\) represents that the electromagnetic field in feedline is at the vacuum and the qubit is at the ground state. Instituting Eqs.~(\ref{6}-\ref{7}) and (\ref{9}-\ref{10}) into Eq.~(\ref{8}), we get:
\begin{eqnarray}
\left\{
\begin{array}{ll}\label{11}
&\omega \phi_{R}(x)=\phi_{R}(x)(-iv_{g}\frac{\partial}{\partial x})+V_1A_{0},\\
&\omega \phi_{L}(x)=\phi_{L}(x)(iv_{g}\frac{\partial}{\partial x})+V_1A_{0},\\
&\omega A_{0}=V_1[\phi_{R}(x)+\phi_{L}(x)]+\omega_{0}A_{0}.\\
\end{array}
\right.
\end{eqnarray}
For the sake of the convenient calculation, we assume that~\cite{2005JS}
\begin{eqnarray}
\left\{
\begin{array}{lll}\label{12}
\phi_{R}(x)&=&e^{ikx}[\theta(-x)+t\theta(x)],\\
\phi_{L}(x)&=&re^{-ikx}\theta(-x),
\end{array}
\right.
\end{eqnarray}
where \(t\)/\(r\) is the transmission/reflection amplitude of the travelling-wave photons. Instituting Eq.~(\ref{12}) into Eq.~(\ref{11}), the transmission amplitude $t$ can be solved as
\begin{equation}
t_{0}(\omega)=\frac{\omega-\omega_{0}}{\omega-\omega_{0}+i\gamma_c},\,
\gamma_c=\frac{V_1^{2}}{v_{g}},
\end{equation}
which is dependent of the frequency of the driven microwave. Consequently, the transmitted spectrum of the microwave photons scattered by the rf-SQUID qubit, without the NMR vibration, can be calculated as
\begin{equation}\label{14}
T_{0}(\omega)=|t_0(\omega)|^2=\frac{(\omega-\omega_0)^2}{v_g^2(\omega-\omega_0)^2+\gamma_c^2}.
\end{equation}
Correspondingly, the phase shift spectrum of the transmitted microwave reads
\begin{equation}\label{15}
\phi_0(\omega)=-\arctan\left(\frac{\gamma_c}{\omega-\omega_{0}}\right).
\end{equation}
\begin{figure}
  \centering
  \includegraphics[width=8.6cm]{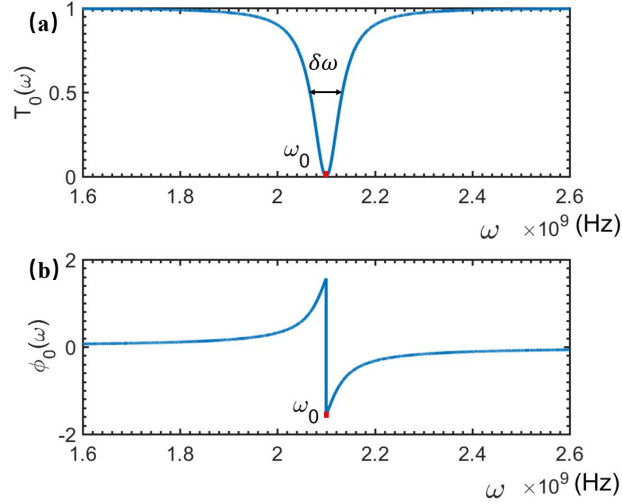}\\
  \caption{The transmitted and phase shift spectra of the travelling microwave scattered simply by the rf-SQUID-based qubit. Here, the relevant parameters are typically set as: $\omega_{0}=2.1\times10^9$Hz, $\gamma_{c}=3.3\times10^7$Hz, and $\delta\omega=6.6\times10^7$Hz.}
  \label{Fig2}
\end{figure}

It is seen from Fig.~\ref{Fig2}(a) that, the transmitted tip is at the microwave-qubit resonant point, i.e., the driving microwave with the frequency $\omega=\omega_0$ is completely reflected. Therefore, by observing the frequency at the dip in the transmitted spectrum of the travelling microwave scattered by the qubit, the eigenfrequency $\omega_0$ of the qubit can be determined. Certainly, due to the coupling dissipation $\gamma_c$, the transmitted dip is not the $\delta$-function. Instead, it shows a Lorentz tip shape with the full width at half minimum (FWHM): $\delta\omega_0=2\gamma_c$. Experimentally, such a FWHM could be served as the uncertainty of the observed dip. In this model, we omitted the internal dissipations of the qubit and treated $\gamma_c$ is the total dissipation of the qubit. On the other hand, Fig.~\ref{Fig2}(b) shows that, if $\omega=\omega_0$ the phase of the reflected microwave is shifted a $\pi$-phase, which is independent of the dissipation of the qubit.

\subsubsection*{Measuring the vibrational frequency of a quantum mechanical NMR}

As shown schematically in Fig.~\ref{Fig1}, if the magnetic field $\bf{B}_0$ is applied, then the vibration of the MNR is excited. Furthermore, let us assume that the vibration of the embedded NMR has been cooled into the quantum regime, i.e., the NMR is treated as a quantum mechanical oscillator (called as the QNMR afterwards) and described by the bosonic operators $\hat{b}$ and $\hat{b}^\dagger$. By a long but direct derivation, the Hamiltonian $\hat{H}_s$ in Eq.~(\ref{eq5}) can be effectively expressed as
\begin{equation}\label{16}
\hat{H}_{q-QNMR}=\omega_{0}|1\rangle\langle 1|+\omega_{b}\hat{b}^{\dagger}\hat{b}+ g_{Q}(\hat{\sigma}_{+}\hat{b}+\hat{\sigma}_{-}\hat{b}^{\dagger}),
\end{equation}
with $g_Q$ being the qubit-QNMR coupling strength (See Methods), $B_0$ is the applied magnetic field in Fig.~\ref{Fig1}, $I_p$ is the amplitude of the circular supercurrent along the rf-SQUID qubit loop, $l$ is the length of the QNMR.

Accordingly,the generic solution of Eq.~(\ref{8}) can be expressed as
\begin{eqnarray}\label{17}
|\psi_{Q}\rangle= {\int dx[\phi _{R}(x)\hat{c}_{R}^{\dagger }(x)+\phi_{L}(x)\hat{c}_{L}^{\dagger }(x)]|\phi_{Q}\rangle}+A_{Q}\hat{\sigma}^{\dagger}|\phi_{Q}\rangle+B_{Q}\hat{b}^{\dagger}|\phi_{Q}\rangle.
\end{eqnarray}
Here, \(|\phi_{Q}\rangle=|0,0,0\rangle\) refers to the scattering ground state, i.e., the electromagnetic field in feedline is at the vacuum, the NMR is at the vibration ground state,
\begin{figure}
  \centering
  \includegraphics[width=8.6cm]{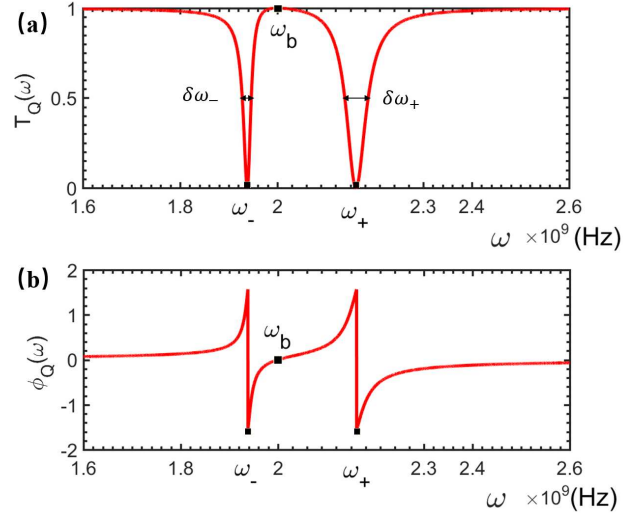}\\
  \caption{The transmitted (a) and phase shift (b) spectra of the travelling microwave scattered by a rf-SQUID-based qubit embedded by a quantum mechanical NMR. The relevant parameters are typically set as: $\omega_{0}=2.1\times10^9$Hz, $\omega_{b}=2.0\times10^9$Hz, $\gamma_{c}=3.3\times10^7$m/s, $g_{Q}=1\times10^8$Hz.}
  \label{Fig3}
\end{figure}
and the qubit is prepared at its ground state $|0\rangle$. Instituting Eqs.~(\ref{6}-\ref{7}) and (\ref{16}-\ref{17}) into Eq.~(\ref{8}), one can easily proof that the coefficients in Eq.~(\ref{17}) are determined by
\begin{eqnarray}
\left\{
\begin{array}{ll}
&\omega \phi_{R}(x)=\phi_{R}(x)(-iv_{g}\frac{\partial}{\partial x})+V_{1}A_{Q},\\
&\omega \phi_{L}(x)=\phi_{L}(x)(iv_{g}\frac{\partial}{\partial x})+V_{1}A_{Q},\\
&\omega A_{Q}=V_{1}[\phi_{R}(x)+\phi_{L}(x)]+\omega_{0}A_{Q}+B_{Q}g_{Q},\\
&\omega B_{Q}=\omega_{b}B_{Q}+g_{Q}A_{Q}.\\
\end{array}
\right.
\end{eqnarray}
With the same method used in the above subsections, we get the transmitted spectrum $T_Q(\omega)=|t_Q(\omega)|^2$, with
\begin{equation}\label{19}
t_{Q}(\omega)=\frac{(\omega-\omega_{b})(\omega-\omega_{0})-g_{Q}^{2}}{(\omega-\omega_{b})(\omega-\omega_{0}+i\gamma_c)-g_{Q}^{2}},
\end{equation}
and also the phase shift spectrum:
\begin{equation}\label{20}
  \phi_{Q}(\omega)=-\arctan[\frac{(\omega-\omega_{b})\gamma_{c}}{(\omega-\omega_{b})(\omega-\omega_{0})-g_{Q}^{2}}],
\end{equation}
of the travelling microwave in the feedline, respectively.

Fig.~\ref{Fig3} shows the calculated transmitted- and phase shift spectra for the case wherein the vibration of the NMR embedded in the qubit is quantum mechanical.

Interestingly, Eqs.~(\ref{19}) and (\ref{20}) imply also that, if the frequency $\omega$ of the travelling microwave is equivalent to $\omega_b$, i.e., the vibrational frequency of the QNMR, then a frequency point with $|t_Q(\omega)|^2=1$ can be observed between two dips in the transmitted spectrum. This indicates that, the frequency $\omega_b$ of the QNMR could be directly determined by observing the frequency point of the travelling microwaves without any reflection, i.e., the frequency of the microwave is completely transmitted without any phase shift.
To measure the vibrational displacement of the QNMR, we need to detect the qubit-QNMR coupling strength $g_Q$. This can be achieved as follows. First, one can see that two transmitted dips centered respectively at $\omega_+$ and $\omega_-$ are observed in the spectrum shown in Fig.~\ref{Fig3}(a). They are determined by solving the equation: $T_Q(\omega)=0$, and read:
\begin{equation}
\omega_{\pm}=\frac{1}{2}[(\omega_{0}+\omega_{b})\pm\sqrt{4g_{Q}^{2}+(\omega_{0}-\omega_{b})^{2}}].
\end{equation}
As a consequence, the qubit-QNMR coupling strength between the qubit and QNMR can be calculated as
\begin{equation}
g_{Q}=\frac{\sqrt{(\omega_{+}-\omega_{-})^{2}-\omega_{0}^2
+2\omega_{0}\omega_{b}-\omega_{b}^{2}}}{2},
\end{equation}
whose estimated accuracy depends on those of the measured frequency points $\omega_b$ and $\omega_{\pm}$. From the simulated spectra shown in Fig.~\ref{Fig3}(a), the frequency uncertainty of the two dips are observed as: $\delta\omega_{-}=1.9\times10^7$Hz for $\omega_{+}=2.162\times10^9$Hz, and $\delta\omega_{+}=4.8\times10^7$Hz for $\omega_{-}=1.938\times10^9$Hz. While, from Eqs.~(\ref{19}) and (\ref{20}), the observed eigenfrequency $\omega_b$ of the NMR could be threaten as the precise value. 

\subsubsection*{Measuring the amplitude (i.e., phonon number) of a quantum mechanical NMR}

Next, with Eq.~(\ref{71}), we know that either the amplitude of the applied magnetic field or the mass of the NMR can be determined as: ${B}_0=g_Q\sqrt{(2m\omega_b)}/(lI_p)$, if $m$ is given; or $m=(B_{0}^2l^2I_p^2)/2g_Q^2\omega_b$, if the $B_0$ is gviven. Physically, by solving the Heisengberg equations: $d\hat{b}/dt=i[\hat{b},\,\hat{H}_1]$ and $d\hat{b}^\dagger/dt=i[\hat{b}^\dagger,\,\hat{H}_1]$, one can get the time-evolution of the Bosonic operators; $\hat{b}(t)$ and $\hat{b}^\dagger(t)$. Then, by using the qubit-QNMR coupling strength $g_Q$ measured above, the mean displacement of the vibrational QNMR can be determined as
\begin{equation}
\bar{z}(t)=\frac{1}{\sqrt{2m\omega_b}}\langle\psi_Q|(\hat{b}(t)+\hat{b}^\dagger(t))|\psi_Q\rangle,
\end{equation}
which is obviously dependent of the quantum state of the system, not only the vibrational quantum state of the QNMR. Therefore, the mean displacement of the QNMR, including the influence from the quantum fluctuation, can be determined. Certainly, such a measurement is a quantum demolition one, as the energy exchange takes place frequently between the qubit and the QNMR. However, following Ref.~\cite{NN2019}, the quantum nondemolition measurement of the quantized vibration of the NMR could be implemented with the present configuration. This can be achieved by adjusting the eigenfrequency $\omega_0$ of the qubit to let the qubit-QNMR work in the dispersive regime, i.e., \(g_Q/\Delta\ll 1\), with \(\Delta=\omega_{0}-\omega_{r}\). under this condition, $\hat{H}_s$ in Eq.~(\ref{eq5}) can be reduced as
\begin{equation}
\hat{H}_{q-QNMR}^{\prime}=\omega_{0}|1\rangle\langle 1|+\omega_{b} \hat{b}^{\dagger} \hat{b}+\frac{ g_{Q}^{2}}{\Delta}\left(\hat{b}^{\dagger} \hat{b}+\frac{1}{2}\right)\hat{\sigma}_{z},
\end{equation}
instead $\hat{H}_{q-QNMR}$ in Eq.~(\ref{16}). Correspondingly, the generic solution
to the Schr\"odinger equation Eq.~(\ref{8}) can be written as
\begin{equation}
|\psi_{QNMR}^{\prime}\rangle=|\psi_{N M R}\rangle \otimes\left[\int d x\sum_{j=L, R}\phi_{j}(x) \hat{c}_{j}^{\dagger}(x)|\phi_{Q}^{\prime}\rangle+A_{e} \hat{\sigma}^{\dagger}|\phi_{Q}^{\prime}\rangle\right],
\end{equation}
with $|\phi_{Q}^{\prime}\rangle=|0,0\rangle$ being the scattered ground state and $|\psi_{QNMR}\rangle$ the vibrational quantized state of the QNMR, which will be nondemolition during \begin{figure}[htbp]
        \centering
        \includegraphics[width=8.6cm]{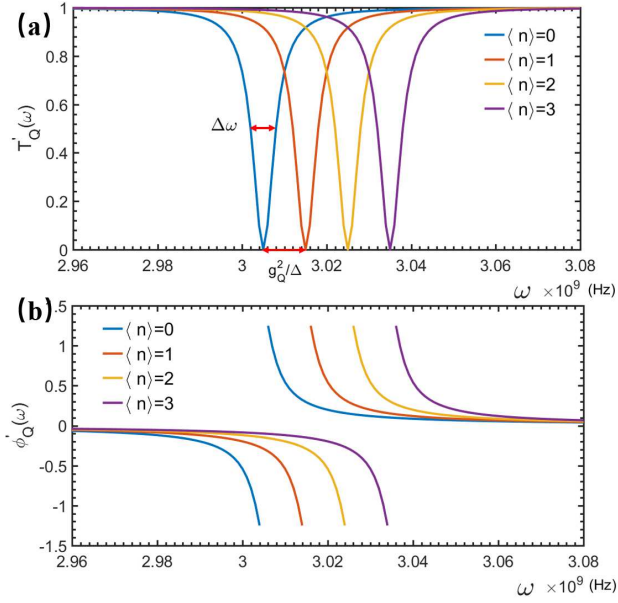}\\
        \caption{(a)Transmitted spectra of  probing wave for  different average phonon numbers \(\langle n\rangle=0,1,2,3,\dots\) in the QNMR. (b)Phase shift of probing wave after  scattering of  qubit dispersively coupled to NMR with  different average photon numbers \(\langle n\rangle=0,1,2,3,\dots\).}
         \label{Fig4}
\end{figure}
the scattering spectral measurements. The coefficients in the above wave function are
determined by
\begin{equation}\label{26}
\left\{\begin{array}{l}
\omega \phi_{R}(x)=-i v_{g} \frac{\partial \phi_{R}(x)}{\partial x}+V_{1} A_{e}, \\
\omega \phi_{L}(x)=i v_{g} \frac{\partial \phi_{L}(x)}{\partial x}-V_{1} A_{e}, \\
\omega A_{e}=V_{1}\left[\phi_{L}(x)-\phi_{R}(x)\right]+A_{e} \frac{g_{Q}^{2}}{2 \Delta}+A_{e} \frac{g_{Q}^{2}}{\Delta}\langle n\rangle,
\end{array}\right.
\end{equation}
where $\langle n\rangle=\langle\psi_{N M R}|\hat{b}^{\dagger} \hat{b}|\psi_{N M R}\rangle$ is
the average phonon number of the vibrational QNMR. Solving Eq.~(\ref{26}) similarly, we get the relevant transmitted- and phase shift spectra:
\begin{equation}
\left|T_{Q}^{\prime}(\omega)\right|=\frac{v_{g}^{2}\left(\omega-\omega_{0}-\frac{g_{Q}^{2}}{2 \Delta}-\frac{g_{Q}^{2}}{\Delta}\langle n\rangle\right)^{2}}{v_{g}^{2}\left(\omega-\omega_{0}-\frac{g_{Q}^{2}}{2 \Delta}-\frac{g_{Q}^{2}}{\Delta}\langle n\rangle\right)^{2}+V_{1}^{4}}
\end{equation}
and
\begin{equation}
\phi_{Q}^{\prime}(\omega)=\arctan \left[\frac{V_{1}^{2}}{v_{g}\left(\omega-\omega_{0}-\frac{g_{Q}^{2}}{2 \Delta}-\frac{g_{Q}^{2}}{\Delta}\langle n\rangle\right)}\right],
\end{equation}
respectively.
It is seen schematically from Fig.~\ref{Fig4} that, the dips in the transmitted spectrum and the phase shifts, near the completely reflected frequency points, are really related to the average number of phonon $\langle n\rangle$ of the QNMR. The centre frequency of the transmission spectrum. The shifted frequency of the dip is dependent on the phonon number of the QNMR. This implies that the average phonon number of the QNMR could be measured by probing how the frequency of the dip is shifted. Again, the FWHM $\Delta\omega=2V_{1}^{2}/v_{g}$ of the observed dip limits the measurement accuracies of the phonon numbers of the QNMR. Experimentally, to differentiate the frequencies corresponding to the nearest two dips induced respectively by the phonon states $|n\rangle$ and $|n+1\rangle$, the condition:
\begin{equation}
V_{1}^{2}<\frac{g_{Q}^{2}}{2 \Delta}v_{g}
\end{equation}
should be satisfied.

\subsubsection*{Measuring the vibrational amplitude of a classical NMR}
Physically, due to the unavoidable dissipation, the quantum feature of the NMR would be lost and the quantum vibration of the NMR becomes the classical one. For the classical NMR (called as the CNMR later) with the amplitude $A_C$, the Hamiltonian $\hat{H}_s$ in Eq.~(\ref{eq5}) reads
\begin{equation}\label{30}
\hat{H}_{q-CNMR}=\omega_{0}|1\rangle\langle 1|+ g_{C}(\hat{\sigma}_{+}e^{-i\omega_bt}+\hat{\sigma}_{-}e^{i\omega_bt}),
\end{equation}
which is simply obtained by replacing the $q$-number Bosonic operators $\hat{b}$ and $\hat{b}^\dagger$ in Eq.~(\ref{16}) as the $c$-number quantities. Here, $g_C=B_{0}l I_p A_C$ is the coupling strength between the qubit and the CNMR and $A_C$ is the amplitude of
the CNMR. In the rotating frame defined by the transformation $U(t)=\exp[i\omega_{b} t\hat{\sigma}_{z}/2]$, the Hamiltonian in Eq.~(\ref{30}) can be rewritten as
\begin{equation}\label{31}
\hat{H}'_{q-CNMR}=(\omega_0+\omega_b)|1\rangle\langle 1|+g_{C}(\hat{\sigma}^{\dagger}+\hat{\sigma}_{-}).
\end{equation}
Here, $\hat{\sigma}^\dagger$ and $\hat{\sigma}_-$ are the Pauli operators of the qubit. The Hamiltonian (\ref{31}) can be easily diagonalized as
\begin{equation}\label{32}
\hat{\tilde{H}}_{q}=\tilde{\omega}_0|\tilde{1}\rangle\langle\tilde{1}|,\,
\tilde{\omega}_0=\sqrt{\frac{(\omega_{0}+\omega_{b})^{2}}{4}+g_{C}^{2}},
\end{equation}
with $\hat{\tilde{\sigma}}_{z}=|\tilde{1}\rangle\langle\tilde{1}|-|\tilde{0}\rangle\langle\tilde{0}|=(\omega_{0}
+\omega_{b})\hat{\sigma}_z/(2\tilde{\omega}_0)+g_C\hat{\sigma}_x/\tilde{\omega}_0$.
Obviously, the embedded CNMR just modifies the eigenfrequency of the qubit without the NMR.
Therefore, replacing just the $\omega_0$, in the Eqs.~(\ref{14}) and (\ref{15}), by $\tilde{\omega}_0$, the transmitted and phase shift spectra of the travelling microwave scattered by the present qubit-CNMR system can be easily expressed as
\begin{equation}
T_{C}(\omega)=|t_C(\omega)|^2=\frac{(\omega-\tilde{\omega}_0)^2}{(\omega-\tilde{\omega}_0)^2+\gamma_c^2}
\end{equation}
and
\begin{equation}
\phi_C(\omega)=-\arctan\left(\frac{\gamma_c}{\omega-\tilde{\omega}_{0}}\right),
\end{equation}
respectively. This indicates that, the spectral shape is the same as the situation that the microwave scattered by the qubit without the NMR, but the resonance point is shifted from $\omega_0$ into $\tilde{\omega}_{0}$.
\begin{figure}
  \centering
  \includegraphics[width=8.6cm]{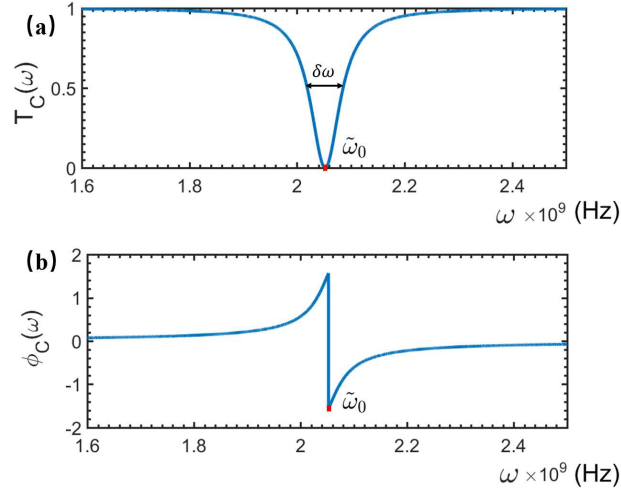}\\
  \caption{The transmitted (a) and phase shift (b) spectra of a rf-SQUID-based qubti embedded by a classical NMR (CNMR). The relevant parameters are set as: $\omega_{0}=2.1\times10^9$Hz, $\omega_{b}=2\times10^9$Hz, $\gamma_{c}=3.3\times10^7$m/s and $g_{C}=1\times10^8$Hz.}
   \label{Fig5}
\end{figure}
Typically, one can see that, differing from the two dips in Fig.~\ref{Fig3}(a) for the qubit-QNMR scattering, Fig.~\ref{Fig5}(a) shows a single dip in the transmitted spectrum of the travelling wave microwave. Therefore, the vibrational feature of the NMR, i.e., either the quantum mechanical or classical vibration, could be calibrated by observing the transmitted spectrum of the travelling microwave scattered by the qubit-NMR; if the observed spectrum has two dips, then the vibration of the NMR is quantum mechanical, while the CNMR refers to one dip spectrum.
Given the vibration frequency $\omega_b$ of the NMR and also the eigenfrequency $\omega_0$ had been measured, the qubit-CNMR coupling strength
\begin{equation}
g_{C}=\sqrt{\tilde{\omega}_0^2-(\omega_0+\omega_b)^2/4},
\end{equation}
can be calculated. Then, the vibrational amplitude of the CNMR can be determined as: $A_{C}= g_C/({\bf B}_0I_{p}l)$. With the set parameters, we get $g_{C}=9.05759\times10^7$Hz, we refer to the following parameters in~\cite{2012J} and assume the
applied magnetic field to be $B_{0}=5$mT, yielding the vibrational amplitude of the CNMR: $A_C\simeq2$nm.

In the above configuration, wherein the travelling microwave is scattered by the rf-SQUID loop qubit, whose coherence might be easily broken by the driving microwave. Alternatively, in the following we consider another configuration, wherein a transmission line resonator (TLR) is introduced to isolate the qubit from the driving microwave and let the later be scattered by the TLR. We shows that, with such an TLR-qubit-NMR configuration, the physical parameters of the NMR can also be measured.

\subsection*{Improving the accuracies with a driven TLR-qubit-NMR system}

\begin{figure}[ht]
        \centering
        \includegraphics[width=8cm]{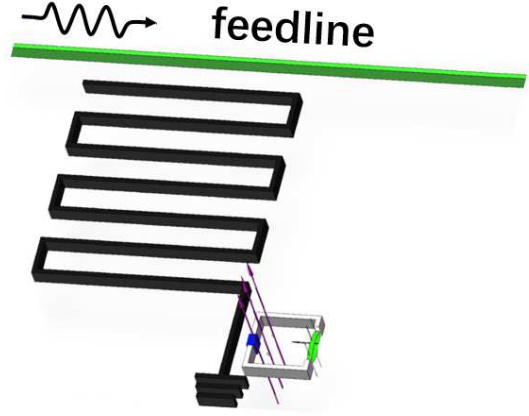}\\
        \caption{Transmitted measurements of the travelling wave scattered by a STLR-qubit-NMR system. Here, the quarter-wavelength STLR is capacitively to the feedline and inductively coupled to the qubit embedded by the NMR.}
         \label{Fig6}
\end{figure}

Following Ref.~\cite{NN2019}, we now consider the configuration shown in Fig.~\ref{Fig6}, wherein the traveling-wave is scattered by the STLR, rather than the qubit. The NMR is still embedded in the rf-SQUID-based qubit.
The Hamiltonian of this
system can be expressed as
\begin{equation}\label{36}
\hat{H}_2=\hat{H}_f+\hat{H}_r+\hat{H}_{fr}+\hat{H}_{rq}+\hat{H}_{s},
\end{equation}
with
\begin{equation}
\hat{H}_r=\omega_{r}\hat{a}_{r}^\dagger \hat{a}_{r}
\end{equation}
describing the fundamental-mode standing-wave photons (with the frequency $\omega_{r}$) in the STLR, $\hat{a}_{r}^\dagger$ and $\hat{a}_{r}$ represent the generation and annihilation operators of the photons in TLR respectively. The interaction between the travelling wave photons transporting along the feedline and the standing-wave photons in the resonator reads
\begin{equation}
\hat{H}_{fr}=\int \delta(x)dx V_{2} \sum_{j=L,R}[\hat{c}_{j}^{\dagger}(x)\hat{a}_{r}+\hat{a}_{r}^{\dagger}\hat{c}_{j}(x)],
\end{equation}
where \(V_2\) is the coupling strength between the travelling wave photons in the feedline and the standing wave photons in the STLR~\cite{gao}.
The coupling between the photons in resonator and the qubit reads
\begin{equation}
\hat{H}_{rq}=g_{rq}(\hat{a}_{r}^{\dagger}\hat{\sigma}_-+\hat{a}_{r}\hat{\sigma}^{\dagger})
\end{equation}
with $g_{rq}$ being the STLR-qubit coupling strength. Similarly, the transmitted spectra of the travelling waves scattered by the present STLR-qubit-NMR system can be calculated by solving the stationary Schr\"odinger equation:
\begin{equation}\label{40}
\hat{H}_2|\tilde\psi\rangle=\omega|\tilde{\psi}\rangle,
\end{equation}
for various vibrational forms of the NMR, i.e., the different forms of the $\hat{H}_s$ shown in Eqs.~(\ref{9},\ref{16},\ref{30}). Noted that, the spectrum of the travelling wave scattered by a single quarter-wavelength STLR had been calculated exactly in Ref.~\cite{gao} and verified by a series of experimental measurements~\cite{Gao01}. Certainly, if the travelling-wave microwave transporting along the feedline is scattered only by the STLR, a single dip centred at $\omega=\omega_r$ could be observed, with the FWHMs $\gamma_r$ being described by the quality factor of the STLR. Next, we will investigate how the spectra are modified by the scatterings of the STLR-qubit-NMR system.

\subsubsection*{The qubit eigenfrequency measurement}

First, if the NMR is absent, i.e., $\hat{H}_s$ in Eq.~(\ref{36}) reduces to $\hat{H}_{q}$ in Eq.~(\ref{9}),
then the generic solution of the Schr\"odinger equation (\ref{40}), with Hamiltonian (\ref{36}), can be written as
\begin{eqnarray}
|\tilde{\psi}_{0}\rangle= {\int dx[\phi_{R}(x)\hat{c}_{R}^{\dagger }(x)+\phi_{L}(x)\hat{c}_{L}^{\dagger }(x)]|\tilde{\phi}_{0}\rangle+C_{0}\hat{\sigma}^{\dagger}|\tilde{\phi}_{0}\rangle}+D_{0}\hat{a}_{r}^{\dagger}|\tilde{\phi}_{0}\rangle.
\end{eqnarray}
Here, \(|\tilde{\phi}_0\rangle=|0,0,0\rangle\) is the ground state of the system with the qubit being at the ground state $|0\rangle$ and the electromagnetic fields in feedline and in TLR at electromagnetic vacuum. The coefficients in the above generic wave function are determined by
\begin{eqnarray}
\left\{
\begin{array}{ll}
&\omega \phi_{R}(x)=\phi_{R}(x)(-iv_{g}\frac{\partial}{\partial x})+V_{2}D_{0},\\
&\omega \phi_{L}(x)=\phi_{L}(x)(iv_{g}\frac{\partial}{\partial x})+V_{2}D_{0},\\
&\omega D_{0}=V_{2}[\phi_{R}(x)+\phi_{L}(x)]+\omega_{r}D_{0}+C_{0}g_{rq},\\
&\omega C_{0}=\omega_{0}C_{0}+g_{rq}D_{0},
\end{array}
\right.
\end{eqnarray}
and can be analytically solved as:
\begin{eqnarray}
\left\{
\begin{array}{lll}
&D_{0}(\omega)=\frac{i v_{g}V_{2} (\omega-\omega_{0})}{iv_{g}(\omega-\omega_{r})(\omega-\omega_{0})-iv_{g}g_{rq}^{2}-V_{2}^{2}(\omega-\omega_{0})},\\
&\tilde{r}_0(\omega)=\frac{V_{2}^{2} (\omega-\omega_{0})}{iv_{g}(\omega-\omega_{r})(\omega-\omega_{0})-iv_{g}g_{rq}^{2}-V_{2}^{2}(\omega-\omega_{0})},\\
&\tilde{t}_{0}(\omega)=\frac{iv_{g}(\omega-\omega_{r})(\omega-\omega_{0})-iv_{g}g_{rq}^{2}}{iv_{g}(\omega-\omega_{r})(\omega-\omega_{0})-iv_{g}g_{rq}^{2}-V_{2}^{2}(\omega-\omega_{0})},\\
&C_{0}(\omega)=\frac{i v_{g}V_{2} g_{rq}}{iv_{g}(\omega-\omega_{r})(\omega-\omega_{0})-iv_{g}g_{rq}^{2}-V_{2}^{2}(\omega-\omega_{0})}.\\
\end{array}
\right.
\end{eqnarray}
As a consequence, the spectra of the transmitted and phase shifted of the travelling microwave can be calculated as
\begin{equation}\label{44}
\tilde{T}_0(\omega)=|\tilde{t}_0(\omega)|^2
=\frac{[v_{g}(\omega-\omega_{r})(\omega-\omega_{0})-v_{g}g_{rq}^{2}]^{2}}{[v_{g}(\omega-\omega_{r})(\omega-\omega_{0})-v_{g}g_{rq}^{2}]^{2}-V_{2}^{4}(\omega-\omega_{0})^{2}}
\end{equation}
 and
\begin{equation}
\tilde{\phi}_0(\omega)=-\arctan[\frac{V_{2}^{2}(\omega-\omega_{0})}{v_{g}(\omega-\omega_{r})(\omega-\omega_{0})-v_{g}g_{rq}^{2}}],
\end{equation}
respectively.
\begin{figure}[ht]
  \centering
  \includegraphics[width=8.6cm]{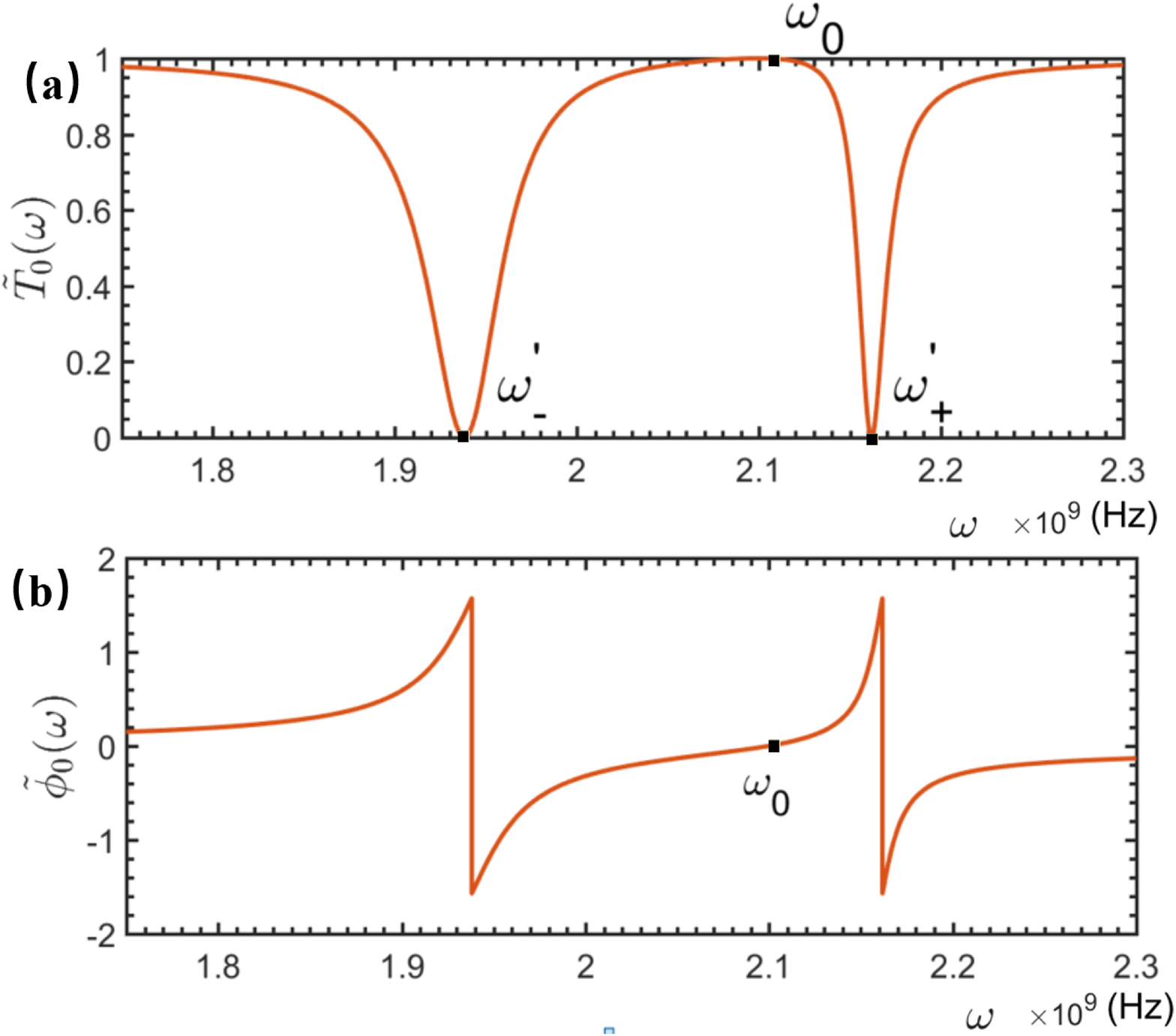}\\
  \caption{The transmitted (a) and phase shift (b) spectra of the travelling microwave scattered by the STLR-qubit system. Here, the relevant paramters are set as: $\omega_{r}=\omega_{b}=2\times10^9$Hz, $v_{g}=3\times10^8$m/s, $\omega_{0}=2.1\times10^9$Hz, $V_{2}=10^8$Hz, $g_{rq}=10^8$Hz. The red and yellow lines refer to without and with the CNMR, respectively.}
   \label{Fig7}
\end{figure}

It is seen from the transmitted spectrum shown in Fig.~\ref{Fig7} that, due to the coupling of the qubit, the single dip centered at $\omega=\omega_0$ in the spectrum of the travelling microwave scattered only by the STLR~\cite{gao} is splitted as the two dips centered at $\omega=\omega'_{\pm}$, with
\begin{equation}
\omega'_{\pm}=\frac{1}{2}(\omega_{0}+\omega_{r}\pm\sqrt{4g_{rq}^{2}+\omega_{0}^{2}-2\omega_{0}\omega_{r}+\omega_{r}^{2}}).
\end{equation}
This is the typical vacuum Rabi splitting phenomenon. Interestingly, in this case, Fig.~\ref{Fig7} and also Eq.~(\ref{44}) indicate that the eigenfrequency $\omega_0$ of the qubit can be determined by observing the frequency point, at which the input microwave is completely transmitted. The observed phenomenon in the spectra, i.e., the frequency at the original completely reflected dip (scattered by a single STLR) is changed as the completely transmitted point of the microwave scattered by the STLR-qubit system. This could be called as the qubit-induced Electromagnetically-induced-transparency (EIT)-like effect of the electromagnetic waves, wherein the qubit is served as the control field to modify the energy structure of the scatter (i.e., the STLR here). Again, by observing the $\omega'_{\pm}$, the STLR-qubit coupling strength can be determined as:
\begin{equation}
g_{rq}=\frac{\sqrt{(\omega'_{+}-\omega'_{-})^{2}-(\omega_{0}-\omega_{r})^{2}}}{2}.
\end{equation}
Typically, if the qubit is resonance with the TLR, then distance between the two dips is $\omega_{+}-\omega_{-}=2g_{rq}$.
\subsubsection*{Vibrational frequency measurement of the QNMR}

Now, let us consider the situation that the vibration of the embedded NMR is quantum mechanical. In this case, the Hamiltonian in Eq.~(\ref{36}) reads $\hat{H}_{q-QNMR}$ shown in Eq.~(\ref{16}). The generic solution to the corresponding Schr\"odinger equation is expressed as
\begin{eqnarray}
|\tilde{\psi}_{Q}\rangle= &{\int dx[\phi_{R}(x)\hat{c}_{R}^{\dagger }(x)+\phi_{L}(x)\hat{c}_{L}^{\dagger }(x)]|\tilde{\phi}_{Q}\rangle+C_{Q}\hat{\sigma}^{\dagger}|\tilde{\phi}_{Q}\rangle} \nonumber\\
&+D_{Q}\hat{a}_{r}^{\dagger}|\tilde{\phi}_{Q}\rangle+E_{Q}\hat{b}^{\dagger}|\tilde{\phi}_{Q}\rangle.
\end{eqnarray}
Here, \(|\tilde{\phi}_Q\rangle=|0,0,0_b,0\rangle\) represents the ground state of the present TLR-qubit-NMR system, which means that the electromagnetic fields in the feedline and the TLR are both in vacuum, the quantized vibration of the NMR is cooled to the vibrational ground state $|0_b\rangle$ and the qubit is prepared at the ground state $|0\rangle$. The coefficients in Eq.~(\ref{40}) are determined by
\begin{eqnarray}
\left\{
\begin{array}{ll}
&\omega \phi_{R}(x)=\phi_{R}(x)(-iv_{g}\frac{\partial}{\partial x})+V_{2}D_{Q},\\
&\omega \phi_{L}(x)=\phi_{L}(x)(iv_{g}\frac{\partial}{\partial x})+V_{2}D_{Q},\\
&\omega D_{Q}=V_{2}[\phi_{R}(x)+\phi_{L}(x)]+\omega_{r}D_{Q}+C_{Q}g_{rq},\\
&\omega C_{Q}=\omega_{0}C_{Q}+g_{rq}D_{Q}+g_{Q}E_{Q},\\
&\omega E_{Q}=\omega_{b}E_{Q}+g_{Q}C_{Q}.
\end{array}
\right.
\end{eqnarray}
They can be analytically solved as:
\begin{eqnarray}
\left\{
\begin{array}{lll}
&D_{Q}(\omega)=\frac{i v_{g}V_{2} A}{iv_{g}(\omega-\omega_{r})A-iv_{g}g_{rq}^{2}-V_{2}^{2}A},\\
&\tilde{r}_{Q}(\omega)=\frac{V_{2}^{2} A}{iv_{g}(\omega-\omega_{r})A-iv_{g}g_{rq}^{2}-V_{2}^{2}A},\\
&\tilde{t}_{Q}(\omega)=\frac{iv_{g}(\omega-\omega_{r})A-iv_{g}g_{rq}^{2}}{iv_{g}(\omega-\omega_{r})A-iv_{g}g_{rq}^{2}-V_{2}^{2}A},\\
&C_{Q}(\omega)=\frac{i v_{g}V_{2} g_{rq}}{iv_{g}(\omega-\omega_{r})A-iv_{g}g_{rq}^{2}-V_{2}^{2}A},\\
&E_{Q}(\omega)=\frac{i v_{g}V_{2} g_{rq} g_{b}}{(iv_{g}(\omega-\omega_{r})A-iv_{g}g_{rq}^{2}-V_{2}^{2}A)(\omega-\omega_{b})},
\end{array}
\right.
\end{eqnarray}
with $A=\omega-\omega_{0}-g_{Q}^{2}/(\omega-\omega_{b})$. Again, the transmitted and phase shifted spectra of the travelling microwave are calculated as
\begin{equation}
\tilde{T}_Q(\omega)=|\tilde{t}_Q(\omega)|^2=\frac{[v_{g}(\omega-\omega_{r})A-v_{g}g_{rq}^{2}]^{2}}{[v_{g}(\omega-\omega_{r})A-v_{g}g_{rq}^{2}]^{2}-V_{2}^{4}A^{2}}
\end{equation}
and
\begin{equation}
\tilde{\phi}_Q(\omega)=-\arctan[\frac{V_{2}^{2}A}{v_{g}(\omega-\omega_{r})A-v_{g}g_{rq}^{2}}],
\end{equation}
respectively.
\begin{figure}[ht]
  \centering
  \includegraphics[width=8.6cm]{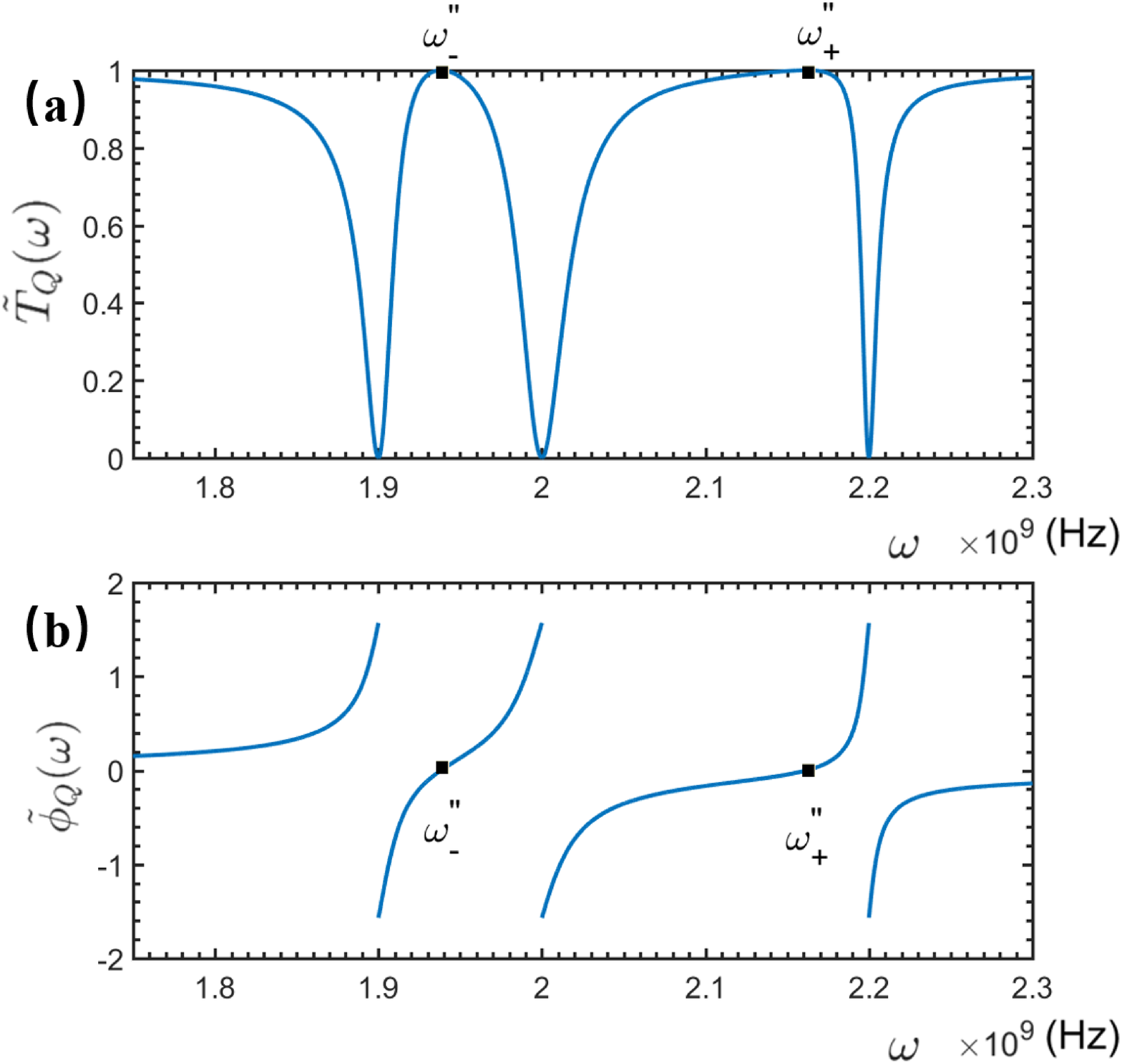}\\
  \caption{The transmitted and phase shift spectra of the travelling microwave scattered by the STLR-qubit-QNMR system. The relevant parameters are set as: $\omega_{r}=\omega_{b}=2\times10^9$Hz, $v_{g}=3\times10^8$m/s, $\omega_{0}=2.1\times10^9$Hz, $V_{2}=10^8$Hz, $g_{rq}=10^8$Hz, and $g_{Q}=1\times10^8$Hz.}
   \label{Fig8}
\end{figure}

One can see from Fig.~\ref{Fig8} that, the quantized vibration of the NMR significantly changes the transmitted and phase shifted spectra of the travelling microwave, typically inducing two EIT-like transparent windows, wherein the travelling microwaves are completely transmitted for \begin{equation}
\omega=\omega^{''}_{\pm}.
\end{equation}
At these frequency points, the phase shifts of the travelling microwave are zero. These behaviors are significantly different from the cases, wherein the vibration of the NMR is absent, and provide more data to measure the physical parameters of the NMR.
For example, with the eigenfrequency $\omega_0$ (the observed completely transmitted frequency $\omega=\omega_0$ in Fig.~\ref{Fig7}) and the two completely transmitted frequency points $\omega^{''}_{-}$ and $\omega^{''}_{+}$ (shown in Fig.~\ref{Fig8}),
\begin{eqnarray}
\omega^{''}_{+}&=\frac{1}{2}(\omega_{0}+\omega_{b}+\sqrt{4g_{b}^{2}+\omega_{0}^{2}-2\omega_{0}\omega_{b}+\omega_{b}^{2}}),\\
\omega^{''}_{-}&=\frac{1}{2}(\omega_{0}+\omega_{b}-\sqrt{4g_{b}^{2}+\omega_{0}^{2}-2\omega_{0}\omega_{b}+\omega_{b}^{2}}).
\end{eqnarray}
Consequently, the vibrational frequency of the NMR can be estimated as
\begin{equation}
\omega_b=\omega^{''}_{+}+\omega^{''}_{-}-\omega_0.
\end{equation}
Furthermore, with the observed frequencies: $\omega_0$ and $\omega^{''}_{\pm}$, shown in Figs.~\ref{Fig7} and \ref{Fig8}, the qubit-QNMR coupling strength could be calculated as:
\begin{equation}
g_{Q}=\sqrt{\omega_{0}(\omega^{''}_{-}+\omega^{''}_{+})-\omega_{0}^{2}
-\omega^{''}_{+}\omega^{''}_{-}}.
\end{equation}
similarly, the vibrational displacement and also the phonon number of the QNMR can be measured.

\subsubsection*{Vibrational amplitude measurement of the CNMR}

If the qubit is embedded by a CNMR, then $\hat{H}_s$ is taken as $\hat{\tilde{H}}_{q}$ in Eq.~(\ref{32}). Consequently, the spectra of the transmitted and phase shifted of the travelling microwave can be calculated as
\begin{equation}
\tilde{T}_C(\omega)=|\tilde{t}_C(\omega)|^2=\frac{[v_{g}(\omega-\omega_{r})B-v_{g}g_{rq}^{2}]^{2}}{[v_{g}(\omega-\omega_{r})B-v_{g}g_{rq}^{2}]^{2}-V_{2}^{4}B^{2}}
\end{equation}
and
\begin{equation}
\tilde{\phi}_C(\omega)=-\arctan[\frac{V_{2}^{2}B}{v_{g}(\omega-\omega_{r})B-v_{g}g_{rq}^{2}}],
\end{equation}
respectively. Above, $B=\omega-\sqrt{((\omega_{0}+\omega_{b})/2)^{2}+g_{C}^{2}}$.
\begin{figure}[ht]
  \centering
  \includegraphics[width=8.6cm]{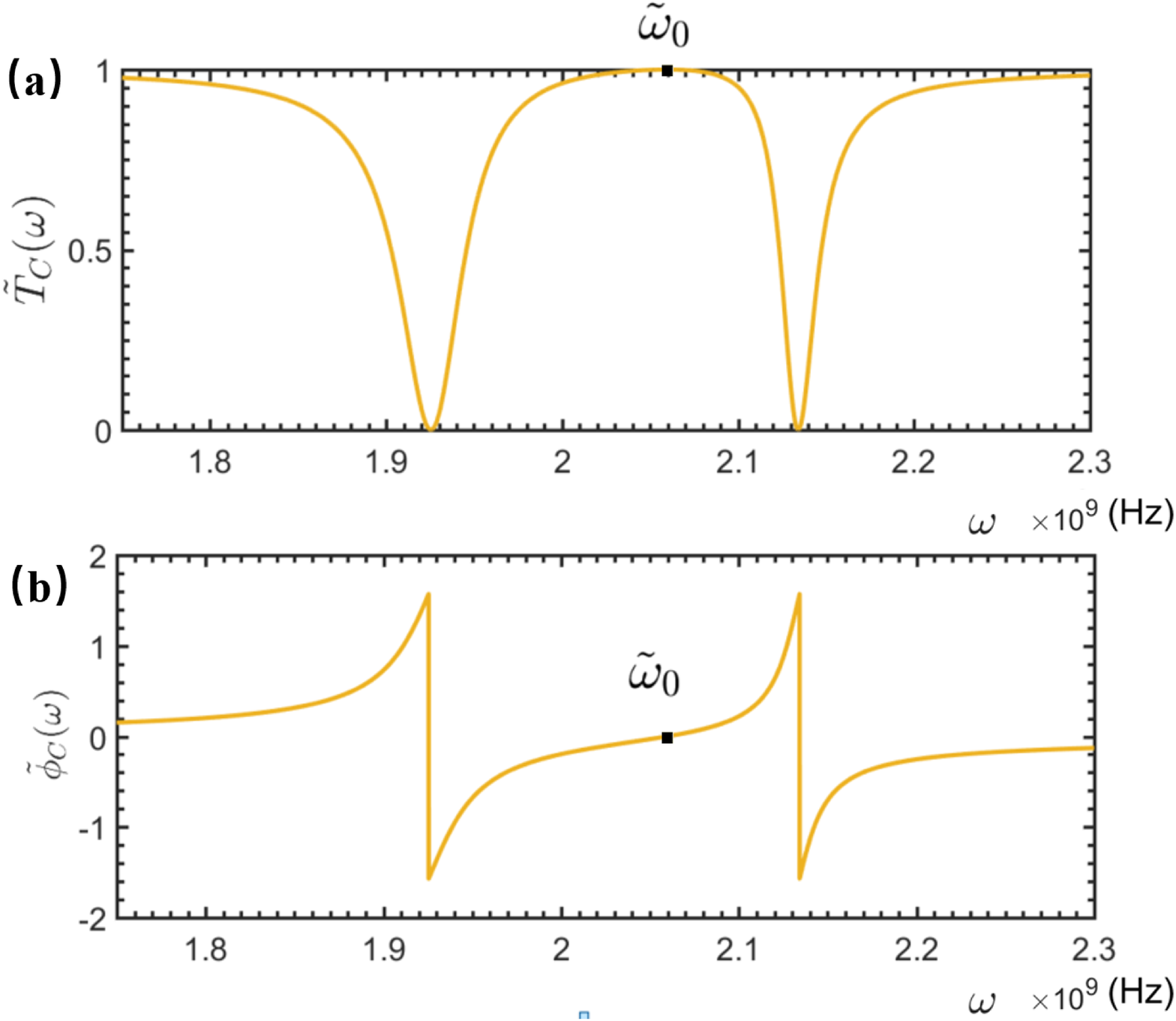}\\
  \caption{The transmitted (a) and phase shift (b) spectra of the travelling microwave scattered by the STLR-qubit-CNMR system. Here, the relevant paramters are set as: $\omega_{r}=\omega_{b}=2\times10^9$Hz, $v_{g}=3\times10^8$m/s, $\omega_{0}=2.1\times10^9$Hz, $V_{2}=10^8$Hz, $g_{rq}=10^8$Hz. The red and yellow lines refer to without and with the CNMR, respectively.}
   \label{Fig9}
\end{figure}
In Fig.~\ref{Fig9} we shows specifically the transmitted and phase shifted spectra of the travelling microwave scatted by the STLR coupled to the qubit embedded by the classical NMR. As mentioned in Fig.~\ref{Fig3}, due to the coupling of the qubit, the transport feature of the travelling microwave scattered by the STLR shows the electromagnetic-induced-transparent (EIT)-like behavior; a transparent window centered at $\omega=\omega_r$ is generated between the two dips (centered at $\omega=\omega_{\pm}$) in the transmitted spectra. In the present case, due to the existence of the classical vibration of the NMR embedded in the qubit, the original EIT-like phenomena (without the NMR) is modified, i.e., the completely transmitted frequency point is shifted into $\omega=\tilde{\omega}_0$, although the width of the transparent window is unchanged. This means that, the modified EIT-like effect can be served as the evidence of the existence of the classical vibration of the NMR.\par
\begin{figure}[ht]
  \centering
  \includegraphics[width=8.6cm]{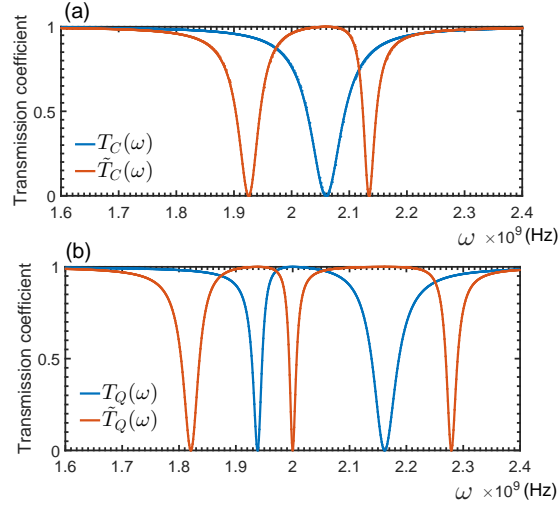}\\
  \caption{Compared the change of transmission peak with or without a resonator, (a)The transmission spectrum contrast of NMR in classical vibration with or without STLR,  (b)The transmission spectrum contrast of NMR in quantum vibration with or without STLR, $\omega_{r}=\omega_{b}=2\times10^9$Hz, $v_{g}=3\times10^8$m/s, $\omega_{0}=2.1\times10^9$Hz, $V_{2}=10^8$Hz, $g_{rq}=10^8$Hz, $g_C=g_{Q}=1\times10^8$Hz.}
   \label{Fig10}
\end{figure}

Fig.~\ref{Fig10} shows the comparison between classical vibration (a) and quantum vibration (b) of the system before and after adding STLR. In general, with the addition of STLR, the trough width of transmission spectrum decreases significantly under the same parameters, which is very important for the accuracy of actual measurement.
\section*{Discussion}

In conclusion, we proposed a spectral measurement method to detect the vibrational and displacement of the NMR, embedded in the rf-SQUID-based qubit. By observing certain specifical frequency points in the measured transmitted and phase shift spectra of the travelling microwave scattered by either the qubit-NMR system or the STLR-qubit-NMR one, we showed that the vibrational frequency and vibrational displacement of the NMR can be measured effectively. By observing the completely transmitted frequency points to determine the qubit-NMR coupling. Interestingly, the proposal provides a quantifiable way to identify the vibrational feature of the NMR, i.e., is the vibration classical or quantum mechanical. In fact, this is the key problem of the NMR being used to implement various precise measurements and quantum information processings.\par
In the present work, we only consider the coupling dissipation and omitted all the internal dissipations of the devices in the system, typically including the qubit, STLR, and the NMR. By simply introducing the relevant non-Hermitian term in the Hamiltonian to phenologically describe the relevant dissipations, we argued that the method proposed here should still be effective. 

\section*{Methods}

\subsection*{The spectral method for a classical harmonic oscillator}

Historically, the stationary spectral method has been widely applied to calibrate the vibrational frequency $\omega_b$ and displacement $z$ of the classical harmonic resonator (HO).
First, by applying a driving force $F(t)$, the equation of the motion of the HO with the internal dissipation $\gamma$ can be expressed as~\cite{U2002}:
\begin{equation}
\ddot{z}+\gamma\dot{z}+\omega_b^2z=\frac{F(t)}{m},
\end{equation}
with $m$, $\omega_b$, and $\gamma$ being the mass, frequency and the dissipative coefficient of the HO, respectively. If the applied force is periodic, i.e., $F(t)=a\cos(\omega_{d} t)$ (with the amplitude $a$ and frequency $\omega_d$), the stationary solution (for $t\gg 1/\gamma$) of the dynamical equation (1) reads: $z(t)=A(\omega_d)\cos[\omega_d t+\phi(\omega_d)]$, with the amplitude- and phase spectra:
\begin{equation}
A(\omega_d)=\frac{a}{m\sqrt{(\omega_{b}^{2}-\omega_{d}^{2})^{2}+\gamma^{2}\omega_{d}^{2}}},\,
\tan[\phi(\omega_d)]=\frac{\gamma\omega_{d}}{\omega_{b}^{2}-\omega_{d}^{2}}.
\end{equation}
As a consequence, by observing the amplitude- and phase spectra under the different frequency driving, the frequency $\omega_b$ of the HO can be determined. Specifically, from the amplitude spectrum, $\omega_b$ can be estimated as the peak value frequency with the accuracy $\delta\omega_b=\gamma$ (i.e, full width at half maximum), which means that the higher precision of the estimation can be obtained for the lower dissipation. While, from the phase spectrum we have $\phi(\omega_{d})=\pi/2$ for $\omega_d=\omega_b$, which is independent of the dissipation of the HO.

Next, two categories: (i) coupling it directly to the sensor~\cite{M2008,CA2008}, and (ii) using the remote radio-wave or optical interferometries~\cite{P2010, D2013,A2018}, are usually applied to detect the vibrational displacement of the HO. However, various unavoidable noises, typically such a the stochastic force $\xi(t)$, limit the sensitivity of these displacement detections. This is because that, in the noise background a dissipative HO is described by the Langiven equation~\cite{R1966}:
\begin{equation}
\frac{d z}{d t}=\upsilon, \frac{d \upsilon}{d t}=-\gamma\upsilon-\omega_{b}^{2} z +\frac{1}{m}\xi(t),
\end{equation}
with $\langle\xi(t)\rangle=0$, but $\langle\xi(t)\xi(t')\rangle\neq 0$. Simply, for the white noise is the Fourier transform of the correlation function of the stochastic force reads: $S_{\xi}(\omega)=\int^{\infty}_{-\infty} d t e^{i \omega t} \langle \xi(t)\xi(0)\rangle=2 m\gamma k_{B}T$. Consequently, the spectrum of the detected displacement is obtained as~\cite{MM2016}
\begin{equation}
z(\omega)=\frac{\xi(\omega)}{m(\omega_{b}^{2}-\omega^{2}-i\gamma\omega)},
\end{equation}
with the spectral density: $S_{x}(\omega)=2 \gamma k_{B}T/\{[m(\omega_{b}^{2}-\omega^{2})^2+\gamma^{2}\omega^{2}]\}$.
Obviously, the reachable sensitivity of the displacement measurement is related to the environment temperature $T$, dissipation parameter $\gamma$ and also the measurement bandwidth.

Physically, although the influence of the noises can be reduced by developing various techniques, typically such as the resonance force microscopy techniques~\cite{SS2009} and the fluid viscosity~\cite{ff2013}, the accuracies of the parameter measurements are very limited as the used probers behave still the classical motions. Basically, the possible improvements should be achieved by using the quantum mechanical probers~\cite{NN2019}. In the following, we discuss how to implement such an improvement by using the qubit and quantum oscillator as the probers.

\subsection*{The derivations of $\hat{H}_s$ for various cases}

In this section, we provide the derivations of $\hat{H}_s$ for various cases, in detail.
\subsubsection*{Hamiltonian of the rf-SQUID based qubit}

For a flux-biased rf-SQUID loop, the Lagrangian can be expressed as
\begin{equation}
\mathcal{L}(\Phi,\dot{\Phi})=\frac{C_{J}}{2}\dot{\Phi}^{2}-\frac{1}{2L}(\Phi-\Phi_{e})^{2}
+\frac{I_{c}\Phi_{0}}{2\pi}\cos(\frac{2\pi\Phi}{\Phi_{0}}),
\end{equation}
where \(\Phi\) and \(L\) are the total flux and inductance of the rf-SQUID loop, respectively. $C_J$ and $I_c$ are the capacitance and critical current of the Josephson junction, respectively. \(\Phi_{e}\) is the biased magnetic flux and \(\Phi_{0}=h/(2e)\) the flux quanta~\cite{XF2017}.
Defining the canonical momenta
\begin{equation}
Q=\frac{\partial \mathcal{L}}{\partial \dot{\Phi}}=C_J\dot{\Phi},
\end{equation}
we have the classical Hamiltonian of the flux-based rf-SQUD loop,
\begin{eqnarray}
\tilde{H}_{0}&=&\dot{\Phi}Q-\mathcal{L}\nonumber\\
&=&\frac{Q^{2}}{2C_{J}}+\frac{1}{2L}(\Phi-\Phi_{e})^{2}-
\frac{I_{c}\Phi_{0}}{2\pi}\cos(\frac{2\pi\Phi}{\Phi_{0}}).
\end{eqnarray}
Formally, this Hamiltonian is equivalent to that for describing the motion of a particle with the "mass" $m=C_J$ moving in the potential:
\begin{equation}
U(\Phi)=\frac{1}{2L}(\Phi-\Phi_{e})^{2}-\frac{I_{c}\Phi_{0}}{2\pi}\cos(\frac{2\pi\Phi}{\Phi_0}).
\end{equation}
Noted that the supercurrent in the loop is determined by $I=-\partial\tilde{H}_{0}/\partial \Phi_{e}=(\Phi-\Phi_{e})/L$~\cite{arXiv}. This means that a clockwise supercurrent (i.e., $I<0$) is generated along the loop if $\Phi_{e}>\Phi$, while for $\Phi_{e}<\Phi$ an anti-clockwise supercurrent (i.e., $I>0$) is generated.
After the usual canonical quantization, the classical Hamiltonian $\tilde{H}_0$ becomes the quantized Hamiltonian:
\begin{equation}
\hat{H}_{0}=\frac{1}{2C_{J}}\frac{\partial^2}{\partial\Phi^2}+\frac{1}{2L}(\Phi-\Phi_{e})^{2}
-\frac{I_{c}\Phi_{0}}{2\pi}\cos(\frac{2\pi\Phi}{\Phi_0}).
\end{equation}

Specifically, for $\Phi_{e}/\Phi_0=0.5$ the potential shows a symmetric double wells around the point $\Phi=\Phi_0$. With the typical parameters:
$\pi I_cL_J/\Phi_0=1, C_J=1.7\times 10^{-14}$F, and $L_J=6\times 10^{-9}$H, the eigenvalue problem of the Hamiltonian $\hat{H}_0$ can be numerically solved and the lowest two eigenvalues are: $E_0=2.7025\times 10^{-23}$J and $E_1=2.7225\times 10^{-23}$J, with the transition frequency between them being $\omega_0=(E_0-E_0)/\hbar=2.1$GHz.
Fig.~11 shows the corresponding eigenfunctions, marked respectively by $|0\rangle$ and $|1\rangle$, of the corresponding eigenvalues $E_0$ and $E_1$.
\begin{figure}[ht]
        \centering
        \includegraphics[width=8.6cm]{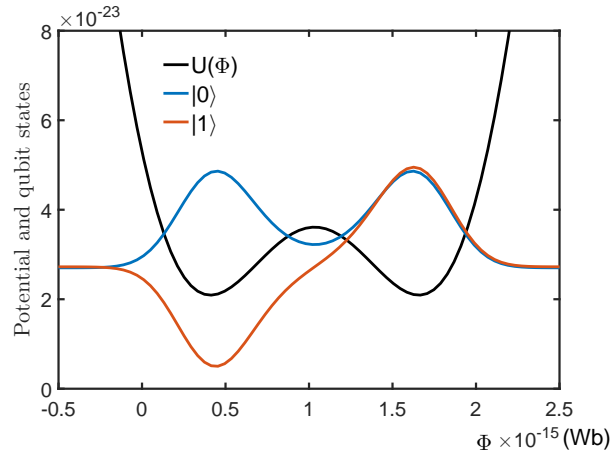}\\
        \caption{Potential function (black line) of a rf-SQUID loop, and the two lowest  energy eigenfunctions (blue- and red lines) of the Hamiltonian $\hat{H}_0$. Here, the parameters are typically set as: \(\Phi_e=0.5\Phi_0\), \(L=6\times10^{-9}\)H, \( I_c=L\Phi_0/\pi\) and \(C_J=1.7\times10^{-14}\)F, respectively.}
         \label{Fig11}
\end{figure}
Physically, a rf-SQUID based qubit can be encoded by the lowest energy eigenstates $|0\rangle$ and $|1\rangle$ of the Hamiltonian $\hat{H}_0$. Its free evolution can be described by the Hamiltonian shown in Eq.~(\ref{9}).
In fact, limited in the subspace of the qubit, we have

\begin{align}
 \hat{H}_{0}&=(|0\rangle\langle 0|+| 1\rangle\langle 1|)\hat{H}_{0}(\mid 0)\langle 0|+| 1\rangle(1 \mid)\nonumber\\
 &=E_{0}|0\rangle\langle 0|+E_{1}|1\rangle\langle 1|+\langle 0|\hat{H}_0|1\rangle|0\rangle\langle 1|
+\langle 1|\hat{H}_0|0\rangle|1\rangle\langle 0| \approx \omega_0|1\rangle\langle 1|=\hat{H}_q.   
\end{align}
This is because that, for the typical parameters: $\langle 0|H_{0}| 1\rangle=\langle 1|H_{0}| 0\rangle=8.02\times10^{-31}$~J, is much less than $\langle 0|\hat{H}_{0}| 0\rangle =E_0=2.7025\times10^{-23}$~J and $\langle 1|\hat{H}_{0}| 1\rangle=2.725\times10^{-23}$~J.
Fig.~\ref{Fig11} shows the distributions of the eigenfunction of the qubit states $|0\rangle$ and $|1\rangle$, respectively. One can easily seen from Fig.~\ref{Fig12} that, the states
\begin{equation}
|L\rangle=\frac{1}{\sqrt{2}}(|0\rangle-|1\rangle),\,\,
|R\rangle=\frac{1}{\sqrt{2}}(|0\rangle+|1\rangle)
\end{equation}
are respectively localized in the symmetric double wells. Therefore, they refer to the clockwise and anti-clockwise supercurrent states~\cite{SN2000,JN2000}, respectively.
\begin{figure}[ht]
        \centering
        \includegraphics[width=8.6cm]{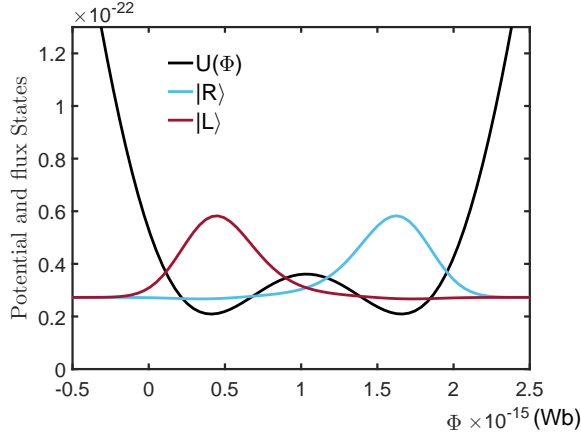}\\
        \caption{The wave functions of the clockwise- (blue line) and anticlockwise (red line) supercurrent states, localized respectively in the left- and right wells of the symmetric potential. The parameters are the same as those used in Fig.~\ref{Fig11}}
         \label{Fig12}
\end{figure}
\subsubsection*{Hamiltonian of the qubit-QNMR system}

Here, we provide the derivation of the Hamiltonians $\hat{H}_{q-QNMR}$ in Eq.~(\ref{16}) and $\hat{H}_{q-CNMR}$ in Eq.~(\ref{30}), respectively.
As shown in Fig.~\ref{Fig1}, the magnetic field $\bf{B}_0$ can be applied to excite the vibrational of the NMR along the $z$-direction. The Lagrangian of the rf-SQUID loop embedded by a NMR can be written as
\begin{equation}
\mathcal{L}(\dot{\Phi},\Phi,\dot{z},z)=\frac{C_{J}}{2}\dot{\Phi}^{2}-U(\Phi)+\frac{1}{2}m\dot{z}^{2}-\frac{1}{2}m\omega_{b}^2z^2-B_{0}Ilz,
\end{equation}
with $m$, $\omega_b$ and $z$ being the mass, vibrational frequency and displacement of the NMR, respeectively. $B_{0}$ is applied along the $y$-direction to provide a restoring force for exciting the mechanical vibration of the NMR. The corresponding Hamiltonian reads:
\begin{equation}
H=\frac{Q^{2}}{2C_{J}}+U(\Phi)+\frac{p^{2}_z}{2m}+\frac{1}{2}m\omega_b^2{z}^{2}+B_{0}Ilz.
\end{equation}

Quantizing the rf-SQUID circuit and the mechanical vibration of the NMR, we have
\begin{equation}\label{70}
\hat{\tilde{H}}_{q-QNMR}=\omega_{0}|1\rangle\langle 1|+\omega_{b}\hat{b}^{\dagger}\hat{b}+\hat{\tilde{H}}_{I},
\end{equation}
with
\begin{equation}\label{71}
\hat{\tilde{H}}_{I}=B_{0}l\sqrt{\frac{1}{2 m\omega_{b}}}(|0\rangle\langle 0|+|1\rangle\langle 1|)\hat{I}(|0\rangle\langle 0|+|1\rangle\langle 1|)(\hat{b}+\hat{b}^{\dagger}).
\end{equation}
Above, the dynamics of the quantized rf-SQUID circuit has limited in the subspace of the rf-SQUID qubit. Also, $\hat{b}$ and $\hat{b}^\dagger$ are the Bosonic operators of the QNMR. Given $U(\Phi)$ is symmetry under the inversion of the magnetic flux around $\Phi=0.5\Phi_{0}$, the values of the diagonal elements: $\langle 0|\hat{I}|0\rangle$ and $\langle 1|\hat{I}|1\rangle$, should be much less than those of the diagonal elements $\langle 1|\hat{I}|0\rangle=\langle 0|\hat{I}|1\rangle$. Here, $\hat{I}=-\partial\hat{H}_0/\partial\Phi_e=(\hat{\Phi}-\Phi_e)/L$. Indeed, with the typical parameters, the numerical results show that:
$I_{p}=\langle 1|\hat{I}|0\rangle=\langle 0|\hat{I}|1\rangle=-9.44\times 10^{-8}$A~\cite{GG2020,316}, and $\langle 0|\hat{I}|0\rangle=2.74\times 10^{-15}$A, $\langle 1|\hat{I}|1\rangle=7.13\times 10^{-13}$A.
As a consequence, Eq.~(\ref{71}) reduces
\begin{equation}
\hat{H}_{I}=g_Q(\hat{\sigma}^\dagger\hat{b}+\hat{\sigma}_-\hat{b}^{\dagger}),\,
g_Q=B_{0}lI_p\sqrt{\frac{1}{2 m\omega_{b}}},
\end{equation}
under the usual rotating-wave approximation (RWA). Therefore, the Hamiltonian $\hat{H}_s$ in Eq.~(\ref{eq5}) becomes that in Eq.~(\ref{16}).

Obviously, if the vibration of the NMR is classical, then the displacement of the CNMR can be represented as a $c$-number (instead the $q$-number): $z=A_{C}\cos(\omega_b t)$ with $A_C$ being the vibrational amplitude. As a consequence, under the RWA the interaction between the qubit and the CNMR can be expressed as:
\begin{equation}
  \hat{H}'_{I}=B_{0}l I_p z (\hat{\sigma}_{+}+\hat{\sigma}_{-})=g_C(\hat{\sigma}_{+}e^{-i\omega_b t}+\hat{\sigma}_{-}e^{i\omega_b t}),
\end{equation}
with $g_C=B_{0}l I_pA_{C}$ being the qubit-CNMR coupling strength. Replacing the $\hat{H}_I$ in Eq.~(\ref{70}) by $\hat{H}'_I$ and removing the pure $c$-number term, Hamiltonian in Eq.~(\ref{70}) becomes that marked as Eq.~(\ref{30}), which describes nothing but the dynamics of the qubit-CNMR system.
 \subsubsection*{The Hamiltonian of the STLR-qubit-NMR system}
 
The Hamiltonian of the standing wave photons in the quarter-wavelength superconducting transmission line resonator (STLR) had been in previous work~\cite{gao}, wherein the mcirowave current near the grounded point in the STLR reads
\begin{equation}
\hat{I}_{r}=\frac{\pi}{2 L_{r}}\sqrt{\frac{1}{\omega_{r} C_{r}}}(\hat{a}_{r}+\hat{a}_{r}^{\dagger}).
\end{equation}
Here, $\omega_r$ is the frequency of the STLR$.\, C_r$ and $L_r$ are the total capacitance and inductance of the resonator, respectively. The quantized Hamiltonian of the rf-SQUID loop, without the NMR, can be expressed as
\begin{equation}
\hat{H}_{0}^{\prime}=\frac{\hat{Q}^{2}}{2C_{J}}+\frac{1}{2 L}(\hat{\Phi}-\Phi_{e}-\hat{\Phi}_{e}^{\prime})^{2}-E_{J} \cos(\frac{2 \pi \hat{\Phi}}{\Phi_{0}})=\hat{H}_0+\hat{H}_{rq}.
\end{equation}
Here, $\hat{H}_0=\omega_0|1\rangle\langle 1|$ being Hamiltonian describing the rf-SQUID based qubit, and
\begin{equation}
\hat{H}_{rq}=M_{rq}\hat{I}\hat{I}_{r}\approx
g_{rq}(\hat{a}_r^\dagger\hat{\sigma}_-+\hat{a}_r\hat{\sigma}^\dagger)
\end{equation}
describing the STLR-qubit interaction, under the RWA approximation. Here, $g_{rq}=\pi M I_p/(2L_r\sqrt{\omega_rC_r})$ is the STLR-qubit coupling strength with $M_{rq}$ the mutual inductance between them.

\section*{Data availability}

All data generated or analysed during this study are included in the submitted article.

\bibliography{sample}

\section*{Acknowledgment}

This work is partially supported by the National Key Research and Development Program of China (NKRDC) under Grant No. 2021YFA0718803, and the National Natural Science Foundation of China (NSFC) under Grant No. 11974290.

\section*{Author contributions statement}

L. F. W. proposed the model, H. Y. G. performs the calculations. Both of them analyzed the results and co wrote the paper. All authors have contributed to the information and material submitted for publication, and all authors have read and approved the manuscript.

\section*{Additional information}

Competing Interests: The authors declare that they have no competing interests.

\end{document}